\documentstyle[12pt]{article}
\begin{document}
\def \inbar{\vrule height1.5ex width.4pt depth0pt}
\def \C{\relax\hbox{\kern.25em$\inbar\kern-.3em{\rm C}$}}
\def \R{\relax{\rm I\kern-.18em R}}
\newcommand{\Z}{\ Z \hspace{-.08in}Z}
\newcommand{\be}{\begin{equation}}
\newcommand{\ee}{\end{equation}}
\newcommand{\bea}{\begin{eqnarray}}
\newcommand{\eea}{\end{eqnarray}}
\newcommand{\p}{\psi}
\renewcommand{\l}{\epsilon}
\newcommand{\f}{\phi}
\newcommand{\g}{\gamma}
\newcommand{\G}{\Gamma}
\newcommand{\e}{\eta}
\newcommand{\z}{\pi}
\newcommand{\m}{\mu}
\newcommand{\n}{\nu}
\newcommand{\s}{\sigma}
\renewcommand{\t}{\tau}
\renewcommand{\a}{\alpha}
\renewcommand{\b}{\beta}
\newcommand{\k}{\kappa}
\renewcommand{\d}{\delta}
\newcommand{\r}{\rho}
\newcommand{\th}{\theta}
\renewcommand{\pt}{\frac{\partial}{\partial t}}
\newcommand{\ppt}{\frac{\partial^{2}}{\partial t^{2}}}
\newcommand{\nn}{\nonumber}
\renewcommand{\ll}{\left[ }
\newcommand{\rr}{\right] }
\newcommand{\kt}{\rangle}
\newcommand{\br}{\langle}
\newcommand{\fs}{\small}
\newcommand{\so}{S_{0}}
\newcommand{\I}{\mbox{1}_{m\times m}}
\newcommand{\In}{\mbox{1}_{n\times n}}
\newcommand{\xo}{x_{0}}
\newcommand{\po}{\psi_{0}}
\newcommand{\eo}{\e_{0}}
\newcommand{\ts}{\tilde{S}}
\newcommand{\pss}{\frac{\partial}{\partial s}}
\newcommand{\tcuf}{\tilde{\cal F}}
\newcommand{\cuf}{{\cal F}}
\newcommand{\oot}{\mbox{\fs$\frac{1}{2}$}}
\newcommand{\iot}{\mbox{\fs$\frac{i}{2}$}}
\newcommand{\cur}{{\cal R}}
\newcommand{\iv}{\imath\! v}
\newcommand{\ib}{\int_{0}^{\b}}
\newcommand{\lll}{\left( }
\newcommand{\rrr}{\right)}
\newcommand{\llc}{\left\{ }
\newcommand{\rrc}{\right\} }
\newcommand{\lpt}{\left.}
\newcommand{\rpt}{\right.}

\title{SUPERSYMMETRY AND THE ATIYAH-SINGER INDEX THEOREM I\hspace{-.4mm}I:
The Scalar Curvature Factor in the Schr\"{o}dinger Equation}
\author{Ali Mostafazadeh \\
\\ Center for Relativity \\ The University
of Texas at Austin \\ Austin, Texas 78712, USA}
\maketitle
\begin{abstract}
The quantization of the superclassical system used in the proof of the
index theorem results in a factor of $\frac{\hbar^{2}}{8}R$ in the
Hamiltonian. The path integral expression for the kernel is analyzed
up to and including 2-loop order. The existence of the scalar curvature
term is confirmed by comparing the linear term in the heat kernel
expansion with the 2-loop order terms in the loop expansion.
\end{abstract}
\newpage

\section{Introduction}
In \cite{a}, a supersymmetric proof of the twisted spin index theorem
is presented. There, the Peierls bracket quantization is applied to the
following supersymmetric Lagrangian:
\bea
L&=&\ll \oot g_{\m\n}\, \dot{x}^{\m}\,\dot{x}^{\n}+\iot g_{\l\g}\,\p^{\l}
\left( \dot{\p}^{\g}+\dot{x}^{\m}\G_{\m\th}^{\g}\p^{\th}\right) \rr +
\label{e1} \\
&&+\k\ll i\e^{a*}\left( \dot{\e}^{a}+\dot{x}^{\s}A_{\s}^{ab}\e^{\b}\right)
+\oot F_{\l\g}^{ab}\,\p^{\l}\p^{\g}\e^{a*}\e^{b}\rr + \nn \\
&&+\mbox{\fs$\frac{\a}{\b}$}\e^{a*}\e^{a} \; .\nn
\eea
The classical ``momenta'' are defined by:
\be
p_{\m}:=g_{\m\n}\dot{x}^{\n}\; .
\label{e2}
\ee
The Peierls bracket quantization leads to the quantization of the
supersymmetric charge:
\be
Q=\mbox{\fs$\frac{1}{\sqrt{\hbar}}$}\p^{\n}\, g^{\frac{1}{4}}\, p_{\n}
g^{-\frac{1}{4}} \; .
\label{e4}
\ee
The quantum mechanical Hamiltonian is then given by:
\be
H=Q^{2}=\oot g^{-\frac{1}{4}}p_{\m}g^{\oot}g^{\m\n}p_{\n}g^{-\frac{1}{4}}
+\mbox{\fs$\frac{\hbar^{2}}{8}$}R-\mbox{\fs$\frac{\k}{2}$}
F_{\l\g}^{ab}\p^{\l}\p^{\g}
\e^{a*}\e^{b} \; .
\label{e6}
\ee
The fact that $Q$ is identified with the (twisted) Dirac operator,
\be
Q \equiv \not\!\! D \; ,
\label{e7}
\ee
defines $H$ uniquely.

Reducing (\ref{e1}) to a purely bosonic system, i.e. setting $\p =\e =\e^{*}
=0$, one arrives at the Lagrangian for a free particle moving on a Riemannian
manifold \footnote{In this paper the case of closed and simply connected
Riemannian manifolds is considered.}.
Equation~(\ref{e6}) is in complete agreement with the analysis of
the reduced system presented in \cite{bd1}. There, the Hamiltonian was defined
by requiring a particular factor ordering, namely by the time ordering
\cite[\S 6.5]{bd1}. Furthermore, it was shown in \cite[\S 6.6]{bd1} that
the term $\frac{\hbar^{2}}{8}R$ in the Hamiltonian contributed to the
kernel:
\be
K(x';t''|x';t') := \br x''=x';t''|x';t'\kt \; ,
\label{e8}
\ee
a factor of
\be
-\mbox{\fs$\frac{i\hbar^{2}}{24}$}\,R(x')\:(t''-t') \; .
\label{e9}
\ee
This was obtained using the heat kernel expansion of (\ref{e8}). The
quantity: (\ref{e9}) is the linear term in the heat kernel expansion.
In \cite[\S 6.6]{bd1}, the 2-loop terms were computed.
It was shown that indeed the loop expansion validates the
existence of $\frac{\hbar^{2}}{8}R$ term. The presence of
the scalar curvature factor in the Schr\"{o}dinger equation is discussed
in \cite{bd2}. For further review see,
\cite{marinov,cecile} and references therein.

The present paper is devoted to the 2-loop analysis of the path integral
formula for the kernel defined by the quantization of (\ref{e1}).
It is shown that indeed the path integral used in the derivation of the index
formula, \cite{a}, corresponds to the Hamiltonian given by (\ref{e4}).
This serves as an important consistency check for the  supersymmetric
proof of the index theorem presented in \cite{a}.

In Section~2, the loop expansion
is reviewed and the relevant 2-loop terms for the system
of (\ref{e1}) are identified. In Sections~3 and 4, the 2-loop
calculations are presented for the spin ($\k =\a =0$) and twisted
spin ($\k =1$) cases, respectively.

To avoid redundancy, the results
of the first part \cite{a} are used freely. As in \cite{a}, the
Latin indices label $\e$'s, the indices from the first and
the second halves of the Greek alphabet label $\p$'s and $x$'s, respectively
\footnote{$\p$'s are labelled by $\g ,\d ,\l ,\th ,\e$.}.
The condensed notation
of \cite{bd1} is also employed. Finally, the following choices are made:
\[ \hbar =1 \; \; ,\; \; \b :=t'' \;\;,\; \mbox{and}\; t'=0 \; .\]

\section{The Loop Expansion}
Let $\Phi^{i}$ denote the coordinate (field) variables of a superclassical
system. Then, if the Lagrangian is quadratic in $\dot{\Phi}$'s, one has
\cite[\S 5]{bd1}:
\be
K(\Phi '',t''|\Phi ',t'):=\br \Phi '',t''|\Phi ',t'\kt = Z
\int_{\Phi ',t'}^{\Phi '',t''}e^{iS[\Phi ]}(sdet\, G^{+}[\Phi ])^{
-\frac{1}{2}}{\cal D}\Phi \; .
\label{e11}
\ee
In the loop expansion of (\ref{e11}), one expands $\Phi$ around the
classical paths $\Phi_{0}$. Defining $\f$ by:
\[ \Phi (t)=:\Phi_{0}(t)+\f (t)\; , \]
\newpage
one obtains:
\bea
\lefteqn{K(\Phi '',t''|\Phi ',t')\, =\, Z\, (sdet\, G_{0}^{+})^{-\frac{1}{2}}
\, e^{i\so}\, \int e^{\frac{i}{2}\f^{i}\, _{i,}S_{0,j}\f^{j}}
\{ \: 1 \: + }  \label{e12} \\
&& \; +\mbox{\fs$\frac{i}{24}$}S_{0,ijkl}\,\f^{l}\f^{k}\f^{j}\f^{i}
-\mbox{\fs$\frac{1}{72}$}S_{0,ijk}\, S_{0,lmn}\, \f^{n}\f^{m}\f^{l}
\f^{k}\f^{j}\f^{i} + \nn \\
&& \;\ll  -\mbox{\fs$\frac{i}{24}$}\left(\epsilon_{1}\, S_{0,ijk}S_{0,lmn}
\, G_{0}^{+nm} + \epsilon_{2}\,S_{0,mni}G_{0}^{+nm}S_{0,jkl}\right)
\f^{l}\f^{k}\f^{j}\f^{i} + \right. \nn \\
&& \;  \left. +\mbox{\fs$\frac{1}{8}$}\epsilon_{3}\, S_{0,jki}G_{0}^{+kj}
S_{0,mnl}G_{0}^{+nm}\, \f^{l}\f^{i} \rr + \cdots \} {\cal D}\f \; ,\nn
\eea
where,
\[ \epsilon_{1}:=(-1)^{m(l+1)+ln}\, ,\;
\epsilon_{2}:=(-1)^{m(i+1)+in}\, ,\mbox{ and}\;
\epsilon_{3}:=(-1)^{j(i+1)+m(l+1)+k+ln}\; . \]
In (\ref{e12}), $G^{+}$ is the advanced Green's function for the
Jacobi operator:
\be
\left( \, _{i,}S_{,j'}\, \right) := \left(
\frac{\stackrel{\rightarrow}{\d}}{\d \Phi^{i}(t)}
S[\Phi ]\frac{\stackrel{\leftarrow}{\d}}{\d \Phi^{j}(t')} \right)\; .
\label{e13}
\ee
The subscript ``$0$'' indicates that the corresponding quantity is
evaluated at the classical path, e.g. $G_{0}^{+ij}:=G^{+ij}[\Phi_{0}]$.
Finally, ``$\cdots$'' are 3 and higher loop order terms.

To evaluate the right hand side of (\ref{e12}), one needs to perform
the following functional Gaussian integrals:
\newcommand{\fgi}{\int e^{\frac{i}{2}\, \f^{i}\, _{i,}S_{0,j}\f^{j}}}
\newcommand{\cud}{{\cal D}\f}
\bea
\fgi \cud &=& c\, (sdet\, G)^{\frac{1}{2}} \nn \\
\fgi \f^{k}\f^{l}\, \cud &=& -ic\, (sdet\, G)^{\frac{1}{2}}\, G^{kl}
\label{e14} \\
\fgi \f^{k}\f^{l}\f^{m}\f^{n}\, \cud &=& (-i)^{2}c\, (sdet\, G)^{\frac{1}{2}}
\, \left( G^{kl}G^{mn}\pm \mbox{permu.} \right) \nn \\
\fgi \f^{k}\f^{l}\f^{m}\f^{n}\f^{p}\f^{q}\, \cud &=& (-i)^{3}c (sdet\, G)^{
\frac{1}{2}}\, \left( G^{kl}G^{mn}G^{pq}\pm \mbox{permu.} \right) \; .\nn
\eea
In (\ref{e14}), ``$G$'' is the Feynman propagator, ``permu.'' are terms
obtained by some permutations of the indices of the previous term,
``$\pm$'' depends on the ``parity'' of $\f$'s appearing on the left
hand side, and ``$c$'' is a possibly infinite constant of functional
integration. The functional integrals in (\ref{e11}) and (\ref{e14})
are taken over all paths with fixed end points:
\[ \f '':=\f (t'')=0 \;\;\mbox{and}\;\;\f ':=\f (t')=0 \; .\]
This justifies the appearence of $G$ in (\ref{e14}).

One must realize that the terms in square bracket in (\ref{e12}) originate
from the expansion of $(sdet\, G^{+})^{-\frac{1}{2}}$ in (\ref{e11}).
For the system under consideration (\ref{e1}), it was shown in
\cite[Sec. 5]{a} that
\be
sdet(G^{+}) = 1 \hspace{2cm}\mbox{for: $x''=x'$}.
\label{e15}
\ee
This simplifies the computations of Sections~3 and 4 considerably.
In view of (\ref{e15}), the square bracket in (\ref{e12}) drops
and (\ref{e12}) reduces to:
\bea
K&=&K_{\rm WKB}\left\{ 1 -\mbox{\fs$\frac{i}{24}$}\, S_{0,ijkl}\left(
G^{lk}G^{ji}\pm \mbox{permu.}\right) + \right. \label{e16} \\
&& \left. +\mbox{\fs$\frac{i}{72}$}\, S_{0,ijk}S_{0,lmn}\left(
G^{nm}G^{lk}G^{ji}\pm \mbox{permu.}\right) +\cdots \right\} \; , \nn
\eea
where,
\be
K_{\rm WKB}:= Zc\, e^{i\so}(sdet\, G)^{\frac{1}{2}}
\label{e17}
\ee
is the ``WKB'' approximation of the kernel. In Sections~3 and~4,
the terms:
\bea
I&:=&S_{0,ijkl}\left( G^{lk}G^{ji}\pm \mbox{permu.}\right) \label{e18} \\
J&:=&S_{0,ijk}S_{0,lmn}\left( G^{nm}G^{lk}G^{ji}\pm\mbox{permu.}\right) \nn
\eea
are computed explicitly. They correspond to the following Feynman
diagrams:
\[ I \equiv \bigcirc \hspace{-1.5mm}\bigcirc \;\;\; \mbox{and}\;\;\;
J \equiv \mbox{\large $\ominus$}\; . \]
\section{2-Loop Calculations for the Case: $\k =\a =0$}
For $\k =\a =0$, the dynamical equations \cite[eq. 28]{a} are
solved by \cite[eq. 90]{a}:
\be
\xo (t) = \xo \;\;\; , \;\;\; \po (t) = \po \; .
\label{e20}
\ee
As in \cite{a}, all the computations will be performed
in a normal coordinate system centered at $\xo$. Since $K$ and $K_{\rm WKB}$
in (\ref{e16}) have the same tensorial properties, the curly bracket in
(\ref{e16}) must be a scalar. This justifies the use of the normal
coordinates.

The Feynman propagator is given by \cite[Sec. 6]{a}:
\bea
G^{\z\xi '}&=&g_{0}^{\z\m}\ll \mbox{\fs$\th (t-t')$} \frac{(e^{\cur (t-\b )}-1)
(1-e^{-\cur t'})}{\cur (1-e^{-\cur \b })}+ \right. \label{e21} \\
&&\hspace{1cm}\left. \mbox{\fs$\th (t'-t)$}
\frac{(e^{\cur t}-1)(1-e^{-\cur (t'-\b)})}{\cur (e^{\cur \b}-1)}\rr_{\m}^{\xi}
\nn \\
G^{\z\d '}&=&G^{\g\xi '}\;=\; 0 \label{e22} \\
G^{\g\d '}&=&\iot g_{0}^{\g\d}\ll \th (t-t')-\th (t'-t) \rr \label{e23}
\eea
where,
\be
\cur = \left( \cur_{\t}^{\m}\right) =\left( g_{0}^{\m\n}\cur_{\t\n}\right)
:= \left( \iot g_{0}^{\m\n}R_{0\t\n\d\th} \po^{\d}\po^{\th}\right)\; .
\label{e24}
\ee

The functional derivatives of the action which appear in (\ref{e18}) are
listed below:
\bea
S_{0,\t\z '\rho ''}&=&\cur_{\t\z ,\rho}\, \mbox{\fs$
\ll\pt \d (t-t')\rr\d (t-t'')$}
+\cur_{\rho\t ,\z}\, \mbox{\fs$\d (t-t')\ll\pt \d (t-t'')\rr$} \nn \\
S_{0,\t\z '\l ''}&=&iR_{0\t\z\g\l}\po^{\g}\, \mbox{\fs$
\ll\pt\d (t-t')\rr \d (t-t'')$}
+i\G_{0,\l\g\t,\z}\po^{\g}\, \mbox{\fs$
\d (t-t')\ll\pt\d (t-t'')\rr$} \nn \\
S_{0,\t\g '\l ''}&=& S_{0,\l\g '\e ''}\: =\: 0 \nn \\
S_{0,\t\z '\rho ''\k '''}&=&\left\{ -g_{0\t\z ,\rho\k}\,
\mbox{\fs$\ll\ppt \d (t-t')\rr \d (t-t'')\d (t-t''')$}+ \right.
\label{e25} \\
&&\hspace{1cm}  -g_{0\t\rho ,\k\z}\, \mbox{\fs$\d (t-t')
\ll\ppt \d (t-t'')\rr \d (t-t''')$} + \nn \\
&&\hspace{2cm} \left. -g_{0\t\k ,\z\rho}\, \mbox{\fs$\d (t-t')
\d (t-t'') \ll\ppt  \d (t-t''')\rr$} \right\}_{1} + \nn \\
&&\left\{ -2\G_{0\t\z\rho ,\k}\, \mbox{\fs$\ll\pt  \d (t-t')\rr
\ll\pt \d (t-t'') \rr \d (t-t''')$} +\right. \nn \\
&&\hspace{1cm} -2\G_{0\t\rho\k ,\z}\, \mbox{\fs$\d (t-t')
\ll\pt  \d (t-t'')\rr \ll\pt \d (t-t''')\rr$} + \nn \\
&&\hspace{2cm}\left. -2\G_{0\t\k\z ,\rho}\,
\mbox{\fs$\ll\pt\d (t-t')\rr\d (t-t'')\ll\pt \d (t-t''')\rr$}\right\}_{2}
+ \nn  \\
&&\left\{ \cur_{\t\z ,\rho\k}\, \mbox{\fs$\ll \pt \d (t-t')\rr \d (t-t'')
\d (t-t''')$} + \right. \nn \\
&&\hspace{1cm}+\cur_{\t\rho ,\k\z}\, \mbox{\fs$\d (t-t')\ll\pt\d (t-t'')\rr
\d (t-t''')$} + \nn \\
&&\hspace{2cm}\left. +\cur_{\t\k ,\z,\rho}\, \mbox{\fs$\d (t-t')\d (t-t'')
\ll\pt  \d (t-t''')\rr$}\right\}_{3} + \nn \\
&& \left\{ i( \G_{0\m\g\t}\G^{0\m}_{\l\z})_{,\rho\k}\po^{\g}\po^{\l}\,
\mbox{\fs$\ll\pt\d (t-t')\rr\d (t-t'') \d (t-t''')$} + \right. \nn \\
&&\hspace{.4cm} +i(\G_{0\m\g\t}\G^{\m}_{0\l\rho})_{,\k\z}\po^{\g}\po^{\l}\,
\mbox{\fs$\d (t-t')\ll\pt\d (t-t'')\rr\d (t-t''')$} + \nn \\
&&\hspace{.8cm}\left. +
i(\G_{0\m\g\t}\G^{\m}_{0\l\k})_{,\z\rho}\po^{\g}\po^{\l}\,
\mbox{\fs$\d (t-t')\d (t-t'')\ll\pt\d(t-t''')\rr$}\right\}_{4} \nn \\
\pagebreak[4]
S_{0,\t\z '\l ''\g '''}&=&iR_{0\t\z\g\l}\,
\mbox{\fs$\ll\pt\d (t-t')\rr \d (t-t'')\d (t-t''')$}+ \nn \\ \nopagebreak[3]
&&\hspace{1cm}+i\G_{0\l\g\t ,\z}\, \mbox{\fs$\d (t-t')
\ll\pt\d (t-t'')\rr\d 9t-t''')$} + \nn \\ \nopagebreak[3]
&&\hspace{2cm}-i\G_{0\g\l\t ,\z}\, \mbox{\fs$
\d (t-t')\d (t-t'')\ll\pt\d (t-t''')\rr$} \nn \\
S_{0,\l\g'\e ''\d '''}&=& 0 \nn
\eea
where the indices are placed on some of the curly brackets for
identification purposes.
$S_{0,\t\z '\rho ''\l '''}$ and $S_{0,\t\l '\g ''\e '''}$ are omitted
because, as will be seen shortly, they do not contribute to (\ref{e18}).

In view of (\ref{e22}), one needs to consider only the following terms:
\bea
I_{1}&:=&S_{0,\t\z'\rho''\k'''}\ll G^{\t\z'}G^{\rho''\k'''}+
G^{\t\rho''}G^{\z'\k'''}+G^{\t\k'''}G^{\z'\rho''} \rr \nn \\
&=&3S_{0,\t\z'\rho''\k'''}G^{\t\z'}G^{\rho''\k'''} \nn \\
I_{2}&:=&S_{0,\t\z'\l''\g'''}\, G^{\t\z'}G^{\l''\g'''} \label{e25.1} \\
J_{1}&:=&S_{0,\t\z'\rho''}\, S_{0,\m'''\n^{\iv}\s^{v}}\, G^{\t\z'}
G^{\rho''\m'''}G^{\n^{\iv}\s^{v}} \nn \\
J_{2}&:=&S_{0,\t\z'\l''}\, S_{0,\m'''\n^{\iv}\g^{v}}\, G^{\t\z'}
G^{\l''\g^{v}}G^{\m'''\n^{\iv}} \; ,\nn
\eea
where,
\be
I = I_{1}+6I_{2}\; .
\label{e26}
\ee
Here, $6$ is a combinatorial factor and $J$ is a linear combination
of $J_{1}$ and $J_{2}$.

\subsection*{Calculation of $I_{1}$ and $I_{2}$}
Let us denote by $I_{1.\a}$ the terms in $I_{1}$ which correspond
to $\{\; \}_{\a}$ in $S_{0,\t\z'\rho''\k'''}$, with $\a =1,2,3,4$.
The computation of $I_{1.\a}$ is in order. One has:
\bea
I_{1.1}&:=& 3\ib dt\ib dt' \lll -g_{0\t\z ,\r\k}\,\ll\ppt\d (t-t')\rr
\, G^{\t\z'}G^{\r\k}\rrr +\nn \\
&&3\ib dt\ib dt'' \lll -g_{0\t\r ,\k\z}\,\ll\ppt\d (t-t'')\rr\,
G^{\t\z}G^{\r''\k}\rrr+\nn \\
&&3\ib dt\ib dt'''\lll -g_{0\t\k ,\z\r}\,
\ll\ppt\d (t-t''')\rr\, G^{\t\z}G^{\r\k'''} \rrr \nn \\
&=&-3\ib dt \lll \ll g_{0\t\z ,\r\k}+2g_{0\z\r ,\t\k}\rr\, G^{\r\k}I^{\t\z}
\rrr \label{e27} \\
&=&3\ib dt \lll R_{0\z\t\r\k}\, G^{\r\k}I^{\t\z} \rrr \nn \\
&=&0\; . \nn
\eea
In (\ref{e27}),
\be
I^{\t\z}:=\ib dt'\,\ll\frac{\partial^{2}}{\partial t^{'2}} \d (t-t')
G^{\t\z'}\rr \; ,
\label{e28}
\ee
and the third and forth equalities are established using \cite[p.
56]{stephani}:
\be
g_{0\m\n ,\s\t}=-\mbox{\fs$\frac{1}{3}$}\lll R_{0\m\s\n\t}+R_{0\n\s\m\t}\rrr
\; ,
\label{e30}
\ee
and
\be
G^{\r\k}=G^{\k\r}\; .
\label{e29}
\ee
Next, we compute:
\bea
I_{1.2}&:=& -6\G_{0\t\z\r ,\k}\ib dt\ib dt'\ib dt'' \lll
\mbox{\fs$\ll\pt\d (t-t')\rr\pt\ll\d (t-t'')\rr$}
\, G^{\t\z'}G^{\r''\k} \rrr + \nn \\
&&-6\G_{0\t\r\k ,\z}\ib dt\ib dt''\ib dt''' \lll
\mbox{\fs$\ll\pt\d (t-t')\rr\pt\ll\d (t-t''')\rr$}
\, G^{\t\z}G^{\r''\k'''} \rrr + \nn \\
&&-6\G_{0\t\k\z ,\r}\ib dt\ib dt'\ib dt''' \lll
\mbox{\fs$\ll\pt\d (t-t')\rr\pt\ll\d (t-t''')\rr$}
 \, G^{\t\z'}G^{\r\k'''}\rrr \nn \\
&=&-12\G_{0\t\z\r ,\k}\ib dt\lll \lpt \frac{\partial}{\partial t'} G^{\t\z'}
\right|_{t'=t}\lpt \frac{\partial}{\partial t''}G^{\r''\k}\right|_{t''=t}
\rrr + \label{e31} \\
&&-6\G_{0\t\r\k ,\z}\ib dt\lll G^{\t\z}\lpt \frac{\partial^{2}}{\partial t''
\partial t'''}G^{\r''\k'''}\right|_{t''=t'''=t} \rrr \; . \nn
\eea
To evaluate the right hand side of (\ref{e31}), one needs the following
relations:
\bea
\lpt \frac{\partial}{\partial t''}G^{\r''\k}\right|_{t''=t}&=&
\lpt \frac{\partial}{\partial t'}G^{\r\k'}\right|_{t'=t}\: =\:
\oot g_{0}^{\r\n}\ll \frac{1+e^{\cur \b}-2e^{\cur t}}{1-e^{\cur\b}}
\rr_{\n}^{\k} \label{e32} \\
\lpt\frac{\partial^{2}}{\partial t'\partial t}G^{\k\r'}\right|_{t'=t}
&=&-\oot g_{0}^{\k\m}\ll\frac{\cur (1+e^{\cur\b})}{
1-e^{\cur\b}}\rr_{\m}^{\r} \; ,\label{e33}
\eea
and \cite[p. 55]{stephani}:
\be
\G_{0\t\z\r ,\k}=-\mbox{\fs$\frac{1}{3}$}\lll R_{0\t\z\r\k}+
R_{0\t\r\z\k}\rrr \; .
\label{e34}
\ee
Equations (\ref{e32}) and (\ref{e33}) are obtained by differentiating
(\ref{e21}), using symmetries of $\cur$ and:
\be
\th (0) := \oot \; .
\label{e35}
\ee
Combinning equations (\ref{e31})-(\ref{e34}) and using (\ref{e21}),
one obtains:
\bea
\lefteqn{I_{1.2}=\lll R_{0\t\z\r\k}+R_{0\t\r\z\k}\rrr \times } \nn \\
\nopagebreak
&&\ib dt \lll
g_{0}^{\z\m}\ll\frac{1+e^{\cur\b}-2e^{\cur t}}{1-e^{\cur\b}}\rr_{\m}^{\t}
g_{0}^{\r\n}\ll\frac{1+e^{\cur\b}-2e^{\cur t}}{1-e^{\cur\b}}\rr_{\n}^{\k}
\rrr + \label{e36} \\ \nopagebreak
&&-\lll R_{0\t\r\k\z}+R_{0\t\k\r\z} \rrr \times  \nn \\ \nopagebreak
&&\ib dt \lll
g_{0}^{\t\n}\ll \frac{e^{\cur\b}-e^{\cur t}-e^{-\cur (t-\b )}+1}{
\cur (1-e^{\cur\b})}\rr_{\n}^{\z}
g_{0}^{\k\m}\ll\frac{\cur (1+e^{\cur\b})}{1-e^{\cur\b}}\rr_{\m}^{\r}
\rrr  \nn
\eea
To identify the terms in (\ref{e36}) which are linear in $\b$, one
may recall that for every integral
\[ {\cal I}(\b ):=\ib dt\, f(\b ,t) \; ,\]
with an analytic integrand in both $t$ and $\b$, the linear term in
$\b$ is given by:
\be
\ll \lpt \frac{\partial}{\partial \b}\right|_{\b =0}{\cal I}(\b )\rr \b
= f(\b =0,t=0)\, \b \; . \label{e37}
\ee
Thus, one needs to examine the integrands in (\ref{e36}). This
leads to
\be
I_{1.2}=\lll R_{0\t\z\r\k}+R_{0\t\r\z\k}\rrr\, g_{0}^{\z\t}g_{0}^{\r\k}\,
\b +O(\b^{2})= R_{0} \b +O(\b^{2})\; ,
\label{e39}
\ee
where, $R_{0}$ is the Ricci scalar curvature evaluated at $\xo$.

Next step is to compute:
\bea
I_{1.3}&:=&3\mbox{\fs$ \lll \cur_{\t\z ,\r\k}\ib dt \, G^{\r\k} \lpt
\frac{\partial}{\partial t'}G^{\t\z'}\right|_{t'=t} +
+\cur_{\t\r ,\k\z}\ib dt\, G^{\t\z} \lpt
\frac{\partial}{\partial t''}G^{\r''\k}\right|_{t''=t} + \rpt $}\nn \\
&&\hspace{1cm}\mbox{\fs$\lpt \cur_{\t\k ,\z\r}\ib dt \, G^{\t\z} \lpt
\frac{\partial}{\partial t'''}G^{\r\k'''}\right|_{t'''=t}\rrr$} \; . \nn
\eea
This is done by using (\ref{e32}) and (\ref{e37}). The result is:
\be
I_{1.3}=O(\b^{2}) \; .
\label{e40}
\ee
The computation of $I_{1.4}$ is similar. Again, one obtains:
\be
I_{1.4}=O(\b^{2}) \; .
\label{e41}
\ee
This completes the calculation of $I_{1}$. Combinning (\ref{e27}),
(\ref{e39}), (\ref{e40}), and (\ref{e41}), one has:
\be
I_{1}=R_{0}\b +O(\b^{2}) \; .
\label{e42}
\ee

The computation of $I_{2}$ is straightforward. Substituting (\ref{e23})
and (\ref{e25}) in (\ref{e25.1}), one has:
\[ I_{2}=\ib dt\ib dt'\ib dt''\ib dt''' \ll {\cal X}_{\t\z\g\l}(t,t',t'',t''')
\: {\cal Y}^{\t\z\g\l}(t,t',t'',t''')\rr \; , \]
where,
\bea
{\cal X}_{\t\z\g\l}&:=& iR_{0\t\z\g\l}
\mbox{\fs$\ll\pt\d (t-t')\rr\d (t-t'')\d (t-t''')$}+ \nn \\
&& +i\G_{0\l\g\t ,\z}\mbox{\fs$\d (t-t')\ll\pt\d (t-t'')\rr\d (t-t''')$}+ \nn
\\
&& -i\G_{0\g\l\t ,\z}\mbox{\fs$\d (t-t')\d (t-t'')\ll\pt\d (t-t''')\rr$} \nn \\
{\cal Y}^{\t\z\g\l}&:=&\mbox{\fs$-\frac{1}{4}$} G^{\t\z'}g_{0}^{\l\g}
\mbox{\fs$\ll \th (t''-t''') -\th (t'''-t'')\rr$} \; . \nn
\eea
Since ${\cal X}$ is antisymmetric in ($\g \leftrightarrow \l$) and
${\cal Y}$ is symmetric in ($\g\leftrightarrow\l$), $I_{2}$ vanishes.
This together with (\ref{e26}) and (\ref{e42}) lead to:
\be
I=R_{0}\b +O(\b^{2}) \; .
\label{e43}
\ee

\subsection*{Calculation of $J_{1}$ and $J_{2}$}
Substituting (\ref{e25}) in (\ref{e25.1}) and peforming the integrations
which involve $\d$-functions, one obtains:
\bea
J_{1}&=&\ib dt\ib dt'''\lll \cur_{\t\z ,\r}\cur_{\m\n ,\s} \,
\lpt \frac{\partial}{\partial t'}G^{\t\z'}\right|_{t'=t}\,
\lpt   \frac{\partial}{\partial t^{\iv}}G^{\n^{\iv}\s^{v}}\right|_{t^{\iv}
=t^{v}=t'''}\, G^{\r\m'''} \rpt +\nn \\
&&+\cur_{\t\z ,\r}\cur_{\s\m ,\n}\,
\lpt \frac{\partial}{\partial t'}G^{\t\z'}\right|_{t'=t}\,
\lpt \frac{\partial}{\partial t^{v}}G^{\n^{\iv}\s^{v}}\right|_{t^{\iv}=
t^{v}=t'''}\, G^{\r\m'''}  +\nn \\
&&+\cur_{\r\t ,\z}\cur_{\m\n ,\s}\,
\lpt \frac{\partial}{\partial t''}G^{\r''\m'''}\right|_{t''=t}\,
\lpt \frac{\partial}{\partial t^{\iv}}G^{\n^{\iv}\s^{v}}\right|_{t^{\iv}=
t^{v}=t'''}\, G^{\t\z}  +\label{e44} \\
&&\lpt +\cur_{\r\t ,\z}\cur_{\s\m ,\n}\,
\lpt \frac{\partial}{\partial t''}G^{\r''\m'''}\right|_{\t''=t}\,
\lpt \frac{\partial}{\partial t^{v}}G^{\n^{\iv}\s^{v}}\right|_{\t^{\iv}=
t^{v}=t'''}\, G^{\t\z} \rrr \nn \\
&=& \ib dt\ib dt'''\lll \cur_{\t\z ,\r}(\cur_{\m\n ,\s}+\cur_{\n\m ,\s})
\ll \cdots \rr +\cur_{\r\t ,\z}(\cur_{\m\n ,\s}+\cur_{\n\m ,\s})
\ll \cdots \rr \rrr \nn \\
&=& 0 \nn
\eea
In (\ref{e44}), the second equality is obtained by rearranging the indices
and using (\ref{e32}). The terms $\ll \cdots\rr$ involve $G$'s and their
time derivatives. The last equality is established using the antisymmetry
of $\cur$:
\be
\cur_{\m\n}=-\cur_{\n\m}\; .
\label{e45}
\ee

The computation of $J_{2}$ is a little more involved. Carrying out the
integrations involving $\d$-functions, one can write $J_{2}$ in the following
form:
\be
J_{2}=\sum_{\a}^{4} J_{2.\a} \; ,
\label{e46}
\ee
where,
\bea
J_{2.1}&:=&-R_{0\t\z\d\l} R_{0\m\n\e\g}\, \po^{\d}\po^{\e} \: \times \nn \\
&&\ib dt \ib dt''' \lll \lpt \lpt
\frac{\partial}{\partial t'}G^{\t\z'}\right|_{t'=t}
\frac{\partial}{\partial t^{\iv}}G^{\m'''\n^{\iv}}\right|_{t^{\iv}=t'''}
G^{\l\g'''} \rrr \label{e47} \\
J_{2.2}&:=&\mbox{\fs$\frac{1}{3}$} R_{0\t\z\d\l}( R_{0\g\e\m\n}+
R_{0\g\m\e\n} ) \po^{\d}\po^{\e} \: \times \nn \\
&&\ib dt \ib dt''' \lll \lpt \lpt
\frac{\partial}{\partial t'}G^{\t\z'}\right|_{t'=t}
\frac{\partial}{\partial t^{v}}G^{\l\g^{v}}\right|_{t^{v}=t'''}
G^{\m'''\n'''} \rrr  \label{e48} \\
J_{2.3}&:=&\mbox{\fs$\frac{1}{3}$} R_{0\m\n\e\g}( R_{0\l\d\t\z}+
R_{0\l\t\d\z} ) \po^{\d}\po^{\e}\: \times \nn \\
&&\ib dt \ib dt''' \lll \lpt \lpt
\frac{\partial}{\partial t''}G^{\l''\g'''}\right|_{t''=t}
\frac{\partial}{\partial t^{\iv}}G^{\m'''\n^{\iv}}\right|_{t^{\iv}=t'''}
G^{\t\z} \rrr \label{e49} \\
J_{2.4}&:=&\mbox{\fs$-\frac{1}{9}$}\ll (R_{0\l\d\t\z}+R_{0\l\t\d\z})
(R_{0\g\e\m\n}+R_{0\g\m\e\n})\rr\, \po^{\d}\po^{\e}\: \times \nn \\
&&\ib dt\ib dt''' \lll
G^{\t\z}G^{\m'''\n'''}\lpt \frac{\partial^{2}}{\partial t^{v}\,
\partial t''}G^{\l''\g^{v}}\right|_{t''=t,t^{v}=t'''} \rrr \; .
\label{e50}
\eea
In (\ref{e48})-(\ref{e50}), use has been made of (\ref{e34}).

Consider the integrals ($:=\int \int$) in (\ref{e47}). $\int \int$ is symmetric
under the exchange of the pairs $(\t ,\z)\leftrightarrow (\m ,\n)$. The
term  $\po^{\d}\po^{\e}$ is antisymmetric in $\d\leftrightarrow\e$.
Thus, the term $(R.R)$ is antisymmetrized in $\l\leftrightarrow\g $.
However, according to (\ref{e23}), $G^{\l\g'''}$ involves $g_{0}^{\l\g}$
which is symmetric in $\l\leftrightarrow\g$. This makes $J_{2.1}$
vanish. Furthermore, using (\ref{e37}) one finds out that
$J_{2.2}$ and $J_{2.3}$ are at least of order $\b^{2}$.
$\int \int$ in (\ref{e50}) is symmetric under
$\t\leftrightarrow\z$, $\m\leftrightarrow\n$, and
$(\m ,\n )\leftrightarrow (\t ,\z )$. This allows only the term
$R_{0\l\t\d\z}R_{0\g\m\e\n}$ to survive in the square bracket in
(\ref{e50}). Moreover, this term is symmetrized in
$(\m ,\n )\leftrightarrow (\t ,\z )$, or alternatively in
$(\l ,\d )\leftrightarrow (\g ,\e )$. Since $\po^{\d}\po^{\e}$ is
antisymmetric in $\d\leftrightarrow\e$,  the surviving
term which is multiplied by $\int\int$ can
be antisymmetrized in $\l\leftrightarrow\g $. However, due to
(\ref{e23}) $\int\int$ is symmetric in these indices. Hence, $J_{2.4}$
vanishes too.

This concludes the computation of the 2-loop terms in the case
$\k =\a =0$. Combinning (\ref{e16}), (\ref{e18}), (\ref{e26}),
(\ref{e43}), (\ref{e44}), and (\ref{e46}),
one finally obtains:
\be
K=K_{\rm WKB} \llc 1 - \mbox{\fs$\frac{i}{24}$} R_{0}\b +O(\b^{2})\rrc\; .
\label{e51}
\ee

\section{2-Loop Calculations for the Case: $\k =1$}
First, the following special case will be considered:
\be
\tilde{K}:=\br x ,\p ,\e^{*}=0 |x,\p ,\e =0 \kt\; .
\label{e52}
\ee
The dynamical equations, \cite[eq. 28]{a}, are solved by \cite[eq.'s 128]{a}:
\bea
\xo (t)&=&\xo \nn \\
\po (t)&=&\po\:=\:\frac{1}{\sqrt{\b}}\tilde{\po} \label{e53} \\
\eo (t)&=&0 \nn\\
\eo^{*}&=&0 \; .\nn
\eea
Following \cite[Sec. 7]{a}, one chooses a normal coordinate system
centered at $\xo$.

The Feynman propagator is given by:
\be
\lll G^{ij'}\rrr=\lll\begin{array}{cccc}
G^{\z\xi'}& 0 & 0 & 0 \\
0 & G^{\g\d'} & 0 & 0 \\
0 & 0 & 0 & G^{ac^{*'}} \\
0 & 0 & G^{a^{*}c'} & 0
\end{array} \rrr\; ,
\label{e54}
\ee
where $G^{\z\xi'}$ and $G^{\g\d'}$ are given by (\ref{e21}) and (\ref{e23}),
respectively. Moreover, one has \cite[eq.'s 135 and 136]{a}:
\be
\begin{array}{ccc}
G^{ac^{*'}}&=&\ll \iot e^{i(\cuf +\frac{\a}{\b}\In )(t-t')}\rr^{ac}
\mbox{\fs$\ll\th (t-t')-\th (t'-t)\rr$} \\
G^{a^{*}c'}&=&\ll \iot e^{i(\cuf^{*}+\frac{\a}{\b}\In )(t-t')}\rr^{ac}
\mbox{\fs$\ll\th (t-t')-\th (t'-t)\rr$} \; .\\
\end{array}
\label{e55}
\ee
Here,
\[ \cuf =\lll \cuf^{ab}\rrr := \lll \oot F_{0\l\g}^{ab}\po^{\l}\po^{\g}
\rrr\; , \]
and $\In$ is the $n\times n$ unit matrix.

The functional derivatives of the action which enter into the computation
of $I$ and $J$, (\ref{e18}), are listed below \footnote{The other terms
are obtained from this list using the rules of changing the order of
differentiation.}:
\bea
\ts_{0,\t\z'\xi''}&=&\lpt S_{0,\t\z'\xi''}\right|_{\k =\a =0}
\label{e56} \\
\ts_{0,\t\z'\l''}&=&\lpt S_{0,\t\z'\l''}\right|_{\k =\a =0}
\label{e57} \\
\ts_{0,\l\g'\th''}&=&0\: =\: \ts_{0,\l\g'\th''\d'''}
\label{e58} \\
\ts_{0,\t\z'\xi''\r'''}&=&\lpt S_{0,\t\z'\xi''\r'''}\right|_{\k =\a =0}
\label{e59} \\
\ts_{0,\t\z'\g''\l'''}&=&\lpt S_{0,\t\z'\g''\l'''}\right|_{\k =\a =0}
\label{e60} \\
\ts_{0,\t\z' c''}&=&0\: =\:\ts_{0,\t\z' c''}
\label{e61} \\
\ts_{0,\t\g' c''}&=&0\: =\:\ts_{0.\t\g' c^{*''}}
\label{e62} \\
S_{,\t a' c''}&=&0\: =\:S_{,\t a^{*'}c^{*''}}
\label{e63} \\
\ts_{0,\t a'c^{*''}}&=&\cuf_{,\t}^{ca}\,\d (t-t')\,\d (t-t'')
\label{e64} \\
\ts_{0,\l\g' c''}&=&0\: =\:\ts_{0,\l\g' c^{*''}}
\label{e66} \\
S_{,\l c'd''}&=&0\: =\: S_{,\l c^{*'}d^{*''}}
\label{e67} \\
\ts_{0,\l c'd^{*''}}&=&F_{0\d\l}^{dc}\po^{\d}\,\d (t-t')\,\d (t-t'')
\label{e68} \\
\ts_{0,\t\z'\xi'' c'''}&=&0\: =\:\ts_{0,\t\z'\xi'' c^{*'''}}
\label{e70} \\
\ts_{0,\t\z'\g'' c'''}&=&0\: =\:\ts_{0,\t\z'\g'' c^{*'''}}
\label{e71} \\
S_{,\t\z' c''d'''}&=&0\: =\:S_{,\t\z' c^{*''}d^{*'''}}
\label{e73} \\
\ts_{0,\t\z' c''d^{*'''}}&=&iF_{0\t\z}^{dc}\,\mbox{\fs$
\ll\pt\d (t-t')\rr\d (t-t'')\d (t-t''')$}+\label{e74} \\
&&\hspace{-1.5cm}
-iA_{0\t ,\z}^{dc}\,\mbox{\fs$\d (t-t')\pt\ll\d (t-t'')\d (t-t''')\rr$}
+\cuf_{,\t\z}^{dc}\,\mbox{\fs$\d (t-t')\d (t-t'')\d (t-t''')$}
\nn \\
\ts_{0,\t\g'\l'' c'''}&=&0\: =\:\ts_{0,\t\g'\l'' c^{*'''}}
\label{e72} \\
\ts_{0,\t\g' c''d^{*'''}}&=&F_{0\l\g,\t}^{dc}\,\d (t-t')\d (t-t'')
\d (t-t''') \label{e75} \\
\ts_{0,\l\g' c''d^{*'''}}&=&F_{0\l\g}^{dc}\, \d (t-t')\d (t-t'')
\d (t-t''')\; . \label{e76}
\eea
Here, `` $\tilde{\;} $ '' 's are placed to indicate that the special case
of (\ref{e52}) is under consideration.

Equations (\ref{e56})-(\ref{e60}), indicate that the 2-loop contributions
due to the terms which involve only the Greek indices are the same as
the case of $\k =\a =0$, i.e. these terms contribute a factor of
$-\frac{i}{24}R_{0}\b$ to the kernel. Consequently, it is sufficient
to show that the remaining 2-loop terms vanish.

\subsection*{Computation of the Terms of Type $I$}
In view of (\ref{e54}) and (\ref{e56})-(\ref{e76}), the following
terms must be considered:
\bea
\tilde{I}_{1}&:=&\ts_{0,\t\z' c''d^{*'''}}\, G^{\t\z'}G^{c''d^{*'''}}
\label{e76.1} \\
\tilde{I}_{2}&:=&\ts_{0,\l\g' c''d^{*'''}}\, G^{\l\g'}G^{c''d^{*'''}}\; .
\label{e76.2}
\eea
Performing the integration overs $\d$-functions, one has:
\bea
\tilde{I}_{1}&=&\ib dt\lll iF_{0\t\z}^{dc}\,\lpt
\frac{\partial}{\partial t'}G^{\t\z'}\right|_{t=t'}G^{cd^{*}} + \rpt
\label{e77} \\
&&\lpt -iA_{0\t ,\z}^{dc}\, G^{\t\z}\ll \lpt
\frac{\partial}{\partial t''}G^{c''d^{*}}\right|_{t''=t} +
\lpt \frac{\partial}{\partial t'''}G^{cd^{*'''}}\right|_{t'''=t}\rr
+\cuf_{,\t\z}^{dc}\, G^{\t\z}G^{cd^{*}} \rrr \; .\nn
\eea
The first and the last terms in the integrand of (\ref{e77}) vanish
because according to (\ref{e55}):
\be
G^{cd^{*}}:=\lpt G^{cd^{*'}}\right|_{t'=t}=0 \; .
\label{e78}
\ee
Moreover, one has:
\be
\lpt \frac{\partial}{\partial t'}G^{c'd^{*}}\right|_{t'=t}=i\d (0)\d^{cd}
=\lpt -\frac{\partial}{\partial t'}G^{cd^{*'}}\right|_{t'=t} \; .
\label{e79}
\ee
Thus, the remaining terms cancel and one obtains:
\[ \tilde{I}_{1}=0 \; .\]
The computation of $\tilde{I}_{2}$ is quite simple. Substituting
(\ref{e23}), (\ref{e55}), (\ref{e76}) in (\ref{e76.2}) and using
(\ref{e78}), one finds:
\be
\tilde{I}_{2}=\ib dt\lll F_{0\l\g}^{dc}\, G^{\l\g}G^{cd^{*}}\rrr =0 \; .
\label{e80}
\ee
The other terms of type $I$ which involve Latin indices are proportional
to $\tilde{I}_{1}$ or $\tilde{I}_{2}$ and hence vanish.

\subsection*{Computation of the Terms of Type $J$}
There are six different terms of type $J$ which must be considered.
These are:
\bea
\tilde{J}_{1}&:=&\ts_{0,\t\z'\xi''}\ts_{0,\k''' c^{\iv}d^{*v}}\,
G^{\t\z'}G^{\xi''\k'''}G^{c^{\iv}d^{*v}} \nn \\
\tilde{J}_{2}&:=&\ts_{0,\t\z'\l''}\ts_{0,\g''' c^{\iv}d^{*v}}\,
G^{\t\z'}G^{\l''\g'''}G^{c^{\iv}d^{*v}} \nn \\
\tilde{J}_{3}&:=&\ts_{0,\t a'b^{*''}}\ts_{0,\z'''c^{\iv}d^{*v}}\,
G^{\t\z'''}G^{a'b^{*''}}G^{c^{\iv}d^{*v}} \nn \\
\tilde{J}_{4}&:=&\ts_{0,\t ab^{*'}}\ts_{0,\z'''c^{\iv}d^{*v}}\,
G^{\t\z'''}G^{a'd^{*v}}G^{c^{\iv}b^{*''}} \nn \\
\tilde{J}_{5}&:=&\ts_{0,\l a'b^{*''}}\ts_{0,\g''' c^{\iv}d^{*v}}\,
G^{\l\g'''}G^{a'b^{*''}}G^{c^{\iv}d^{*v}} \nn \\
\tilde{J}_{6}&:=&\ts_{0,\l a'b^{*''}}\ts_{0,\g'''c^{\iv}d^{*v}}\,
G^{\l\g'''}G^{a'd^{*v}}G^{c^{\iv}b^{*''}} \; .\nn
\eea
The following relations are useful in the computation of $\tilde{J}$'s.
Using (\ref{e78}), one has:
\bea
\ts_{0,\k c'd^{*''}}\, G^{c'd^{*''}}&=&\cuf_{,\k}^{dc}\, G^{cd^{*}}\: = \:
0 \label{e81} \\
\ts_{0,\g c'd^{*''}}\, G^{c'd^{*''}}&=&F_{0\d\g}^{dc}\po^{\d}\, G^{cd^{*}}
\: = \; 0  \; .\label{e82}
\eea
Equations (\ref{e81}) and (\ref{e82}) lead immidiately to:
\be
\tilde{J}_{\a}=0 \;\;\;\;{\rm for:}\; \a =1,2,3,5\; .
\label{e83}
\ee
It remains to calculate $\tilde{J}_{4}$ and $\tilde{J}_{6}$. In view
of (\ref{e64}) and (\ref{e68}), one has:
\bea
\tilde{J}_{4}&=&\cuf_{,\t}^{ba}\cuf_{,\z}^{dc}\,\ib dt\ib dt'''\,
G^{\t\z'''}G^{ad^{*'''}}G^{c'''b^{*}}
\label{e84} \\
\tilde{J}_{6}&=&F_{0\d\l}^{ba}F_{0\th\g}^{dc}\po^{\d}\po^{\th}\,
\ib dt\ib dt'''\, G^{\l\g'''}G^{ad^{*'''}}G^{c'''b^{*}} \; ,
\label{e85}
\eea
Using (\ref{e37}) and examining the integrands in (\ref{e84})
and (\ref{e85}), one finally arrives at:
\be
\tilde{J}_{\a}=O(\b^{2}) \;\;\;\;\;{\rm for:}\; \a =4,6 \; .
\label{e86}
\ee

This concludes the 2-loop calculations for the special case of (\ref{e53}).
For this case the kernel is given by:
\be
\tilde{K}=\tilde{K}_{\rm WKB}\lll 1 -\mbox{\fs$\frac{i}{24}$}R_{0}\b
+ O(\b^{2}) \rrr \; .
\label{e87}
\ee
In the rest of this section, it is shown that the same conclusion
is reached even for the general case where $\e \neq 0 \neq \e^{*}$,
i.e. for
\[ K := \br x,\p ,\e^{*}|x,\p ,\e \kt \; .\]

In the general case, the dynamical equations \cite[eq. 28]{a} are
solved by \\ \cite[eq.'s 123, 153, \& 154]{a}:
\bea
\xo (t)&=&\xo + O(\b ) \nn \\
\po (t)&=&\frac{1}{\sqrt{\b}}\ll \tilde{\p}_{0}+O(\b )\rr
\label{e88} \\
\eo^{a}(t)&=&O(1)\nn \\
\eo^{a*}(t)&=&O(1)\; .\nn
\eea
It is easy to check that the terms in $S_{0,\cdots}$'s which
involve $\e$'s or $\e^{*}$'s, and thus survive in the general case,
are all of higher order in $\b$ than the terms considered above.
Therefore, the contribution of these terms are of order $\b^{2}$ or
higher. For instance,
\[
S_{0,\t\z'\r''}=\ts_{0,\t\z'\r''}+\Sigma_{\t\z\r} \]
where,
\[
\ts_{0,\t\z'\r''}=O(\b^{-4})\; \]
is given by (\ref{e56}), and
\bea
\Sigma_{\t\z\r}&:=&\ll i(A_{0\z ,\t}^{ab}-A_{0\t ,\z}^{ab})_{,r}
\eo^{a*}\eo^{b}+
-iA_{0\t ,\z\r}^{cd}(\dot{\e}_{0}^{c*}\eo^{d}+\eo^{c*}\dot{\e}_{0}^{d})+
\rpt \nn \\
&&+\lpt i(A_{0\s ,\t}^{cd}-A_{0\t ,\s}^{cd})_{,\z\r}\dot{\xo}^{\s}
\eo^{c*}\eo^{d}+\cuf_{,\t\z\r}\eo^{c*}\eo^{d}\rr\d (t-t')\d (t-t'') +
\nn \\
&&+i(A_{0\r ,\t}-A_{0\t ,\r})_{,\z}\d (t-t')\frac{\partial}{\partial t}
\ll \d (t-t'')\rr \nn \\
&=& O(\b^{-3})\; . \nn
\eea
To determine the order of the terms in $\b$, one proceeds as in
\cite[Sec. 7]{a}, namely one makes the following change of time
variable:
\be
s:=\frac{t}{\b}\, \in [0,1] \; .
\label{e89}
\ee
For example, one has:
\[ \frac{\partial}{\partial t}\d (t-t') =\b^{-2}\frac{\partial}{\partial s}
\d (s-s')=O(\b^{-2}) \; . \]

The terms for which the above argument might not apply are those
that vanish identically in the special case of (\ref{e53}) but
survive otherwise. These are:
\bea
J_{7}&:=&S_{0,\t\z' c''}S_{0,\xi'''\r^{\iv}d^{*v}}\,
G^{\t\z'}G^{\xi''\r^{\iv}}G^{c''d^{*v}} \nn \\
J_{8}&:=&S_{0,\t\z' c''}S_{0,\xi'''\r^{\iv}d^{*v}}\,
G^{\t\xi'''}G^{\z'\r^{\iv}}G^{c''d^{*v}} \nn \\
J_{9}&:=&S_{0,\t\g' c''}S_{0,\z'''\l^{\iv}d^{*v}}\,
G^{\t\z'''}G^{\g'\l^{\iv}}G^{c''d^{*v}} \nn \\
J_{10}&:=&S_{0,\l\g' c''}S_{0,\d'''\th^{\iv}d^{*v}}\,
G^{\l\g'}G^{\d'''\th^{\iv}}G^{c''d^{*v}} \nn \\
J_{11}&:=&S_{0,\l\g' c''}S_{0,\d'''\th^{\iv}d^{*v}}\,
G^{\l\d'''}G^{\g'\th^{\iv}}G^{c''d^{*v}}\; . \nn
\eea
However, one can easily show that the contribution of these terms to the
kernel is of the order $\b^{2}$. The following relations are
useful:
\bea
S_{0,\t\z' c''}&=&O(\b^{-3})\: =\: S_{0,\t\z' c^{*''}} \nn \\
S_{0,\t\g' c''}&=&O(\b^{-\frac{5}{2}}) \nn \\
S_{0,\l\g' c''}&=&O(\b^{-2})\: =\: S_{0,\l\g' c^{*''}} \nn \\
G^{\t\z'}&=&O(\b ) \nn \\
G^{\g\l'}&=&O(1)\: =\:G^{cd^{*'}}\; . \nn
\eea
Furthermore, each integral:
\[
\ib dt\cdots =\b \int_{0}^{1} ds \cdots \]
contributes a factor of $\b$. Putting all this together, one finds out
that $J_{7},\cdots ,J_{11}$ are all of order $\b^{2}$.

This concludes the 2-loop calculations for the general case of
(\ref{e88}). The final result is:
\be
K=K_{\rm WKB}\ll 1-\mbox{\fs$\frac{i}{24}$}R_{0}\b +O(\b^{2})\rr\; .
\label{e90}
\ee
Equation (\ref{e90}) verifies the existence of the scalar curvature
factor in the Hamiltonian and provides a consistency check for the
supersymmetric proof of the Atiyah-Singer index theorem presented
in \cite{a}.

It must be emphasized that the 2-loop term in (\ref{e90})
does not contribute to the index formula \cite[eq. 147]{a}.
This is simply because the $\p$-integrations in \cite[eq. 87]{a}
eliminate such a term.

\section{Conclusion}
The scalar curvature factor in the Schr\"{o}dinger equation
yields a factor of $-\frac{i}{24}R\b$ in the heat kernel
expansion of the path integral. The loop expansion provides an
independent test of the validity of this assertion. In particular,
this factor is obtained in the 2-loop order. Thus, it is shown
that indeed the path integral used in the derivation of the index
formula corresponds to the Hamiltonian defined by the twisted
Dirac operator.

\section*{Acknowledgements}
The author would like to thank Professor Bryce DeWitt who
had originally suggested this project. He has also brought
the author's attention to an algebraic mistake in the first
draft of this paper. The author would also like to express
his gratitute to Professor Cecile DeWitt-Morette for her
carefully reading the second draft and making many constructive
comments and suggestions.

\end{document}